\documentclass[review]{elsarticle}

\usepackage{graphicx}
\usepackage{amsmath}
\usepackage[utf8]{inputenc}
\DeclareUnicodeCharacter{200B}{-}
\begin{document}

\begin{frontmatter}

\title{Meromorphic Solutions of Modified Quintic Complex Ginzburg-Landau Equation}

\author[undana1,ictmp]{Herry F. Lalus\footnote{email : herrylalus@staf.undana.ac.id}}
\author[teori,ictmp]{W. Hidayat \footnote{email : wahid@fi.itb.ac.id}}
\author[undana2]{Meksianis Z. Ndii\footnote{email : meksianis.ndii@staf.undana.ac.id}}
\author[teori,ictmp]{F. P. Zen\footnote{email : fpzen@fi.itb.ac.id}}

\address[undana1]{Department of Physics University of Nusa Cendana, Kupang-NTT, Indonesia}
\address[undana2]{Department of Mathematics University of Nusa Cendana, Kupang-NTT, Indonesia}
\address[teori]{Theoretical High Energy Physics and Instrumentations Research Division, Faculty of Mathematics and Natural Sciences, Institut Teknologi Bandung, Bandung 40132, Indonesia.}
\address[ictmp]{Indonesia Center for Theoretical and Mathematical Physics (ICTMP), Bandung, Indonesia.}  

\begin{abstract}
	In this paper, the meromorphic solution of the modified quintic complex Ginzburg-Landau equation (CGLE) is analysed.  We found the general explicit solutions to the equation in three different forms, yield simply periodic, doubly periodic and rational solution. Firstly, this equation was transformed to nonlinear ordinary differential equation and then we solved it by using a powerful algorithm proposed by Demina and Kudryashov, based on the existence of Laurent series. Finally, we have the meromorphic solution of the equation, and to verify these solutions, we showed a special case which we constructed from the general form. 
\end{abstract}


\end{frontmatter}


\section{Introduction}
Nonlinear partial differential equations have been applied in many areas including nonlinear physics. The main challenge is to find the analytical solutions of nonlinear partial differential equations. A general approach for obtaining analytical solutions is to transform the nonlinear partial differential equations into nonlinear ordinary differential equations. 
A number of research have been conducted to find exact solutions of nonlinear partial differential equations\cite{sulaeman2015,sulaeman2012,sulaeman2012b}. 


In recent years there have been many papers that mention about the exact solutions of various nonlinear differential equations using various methods. The results in these papers they generally refer to them as new solutions of various equations the nonlinear differentials studied. However, against these newly perceived solutions, finally by Kudryashov and some writers \cite{Kud_Mer_Sol_N_ODE,Kud_Be_Careful_Exp_M,Kud_A_Note_on_Kawa_Exp,Kud_Comenton_FisherEq_2009,Kud_Logui_Be_Careful_2009,Kud_Seven_Common_2009,Kud_Sinelshchikov_A_Note_JMeq_2010,Kud_n_Sinel_A_Note_FifthO_KdV_Exp,Park_A_Note_on_Seventh_o_KdV} have shown that most
of these solutions are "well-known" solutions. Kudryashov pointed out that "new solutions" as in \cite{Assas,Zheng2009,Salas} and many other papers, are identical forms with others and is only distinguished by the expression of trigonometric identities, hyperbolic functions, and constants, which actually come from the same Laurent series. For example, Salaz \textit {et al.} \cite {Salas} showed that they found nine new solutions of the Burger equation. However, Kudryashov \cite{Kud_Mer_Sol_N_ODE} showed that solutions are identical and are only distinguished by trigonometric identities, hyperbolic functions, and constants. Furthermore, the solutions are derived from Laurent series of Riccati Equations, which has been changed from Burger's Equation. We quote an interesting statement from Kudrashov \cite {Kud_Mer_Sol_N_ODE} that "We will illustrate the exact solution of the nonlinear equations in the equations determined by the Laurent series for nonlinear solutions on differential equations." Therefore, if we aim to find a general solution of nonlinear differential equations, then the first step is to look for the existence of the Laurent series from the equation. This is a condition that must be fulfilled.

A novel algorithm  to construct explicit meromorphic solution for autonomous nonlinear ordinary differential equations has been proposed by  Demina and Kudryasov \cite{Demina_Kud_Explicit_Expression}. This method is built on the existence of Laurent series of differential equations studied. This method is very powerful to find the general solutions of nonlinear differential equations. By using this algorithm, we can find analytical solutions in three different forms, which cannot be found using the other methods. The other methods can only find a special case of the general solutions of using Demina and Kudryashov approach. An important point of a good method is that the method provides space for each solution obtained using the method to be re-entered into the studied equation, in order to verify the correctness of the results obtained. This method expressly requires that any solution obtained using the algorithm must be verified in this way.  Details of the algorithm can be found in \cite{Demina_Kud_Explicit_Expression, Demina_Kud_From_Lau_Kawa}.

The aim of this paper is to analyze the following equation 
\begin{equation}	
\label{Pers_GL} i\frac{\partial \psi}{\partial t}=P\frac{{\partial^2\psi}}{\partial x^2}+\gamma \psi+Q_1|\psi|^4\psi+iQ_2|\psi|^2\frac{\partial \psi}{\partial x}+iQ_3\psi^2\frac{\partial \psi^*}{\partial x},
\end{equation}
where the coefficients $P$, $Q_1$, $Q_2$, $Q_3$ and $\gamma$ are real and are physical parameters.
Equation (\ref{Pers_GL}) is called the Derivative Nonlinear Schrodinger (DNLS) equation with potential term, or the modified quintic complex Ginzburg-Landau equation (CGLE).

Equation (\ref{Pers_GL}) can be found in several problems in physics such as transmission line or as wave propagation on a discrete nonlinear transmission line. Kengne \textit {et al.} ~\cite{Kengne2006} attempted to solve this equation but they can not find the general solution of this equation. In parts A and B of their work, they always use ansatz for their solution. This can not be the foundation for ensuring that the solution is common. Therefore, a more fundamental approach is needed to determine the general solution, and that approach is the algorithm mentioned earlier.

The solutions discovered by Kengne and Liu have also been proven by Nickel and Schurmaan in \cite{Kengne2007} that such solutions are not common  (based on the solution by Whittaker and Watson \cite{Whittaker_Whatson_A_Course}).
They showed in detail that all solutions \cite{Kengne2006} are merely special cases of the general solutions that they describe in \cite{Kengne2007}.

However, the general solution stated in \cite{Kengne2007} is still in an implicit form. This can be seen from equation (2) in \cite{Kengne2007}. The nonlinear ordinary differential equation (transformed from the quintic Ginzburg-Landau  equation) is still defined again with a new function $R(w)$, so in the end, the solution formulation generally contains the function. Actually, the definition of this new function is not necessary if we use the Demina-Kudryashov algorithm. In this algorithm, we directly solve the ordinary nonlinear differential equation without using a new function definition. Consequently, the solution we get really only contains all the variables and constants that play a role in the equation.

The solution equation analysis (\ref{Pers_GL}) begins by transforming the equation to a nonlinear ordinary differential equation. Then, we use the algorithm proposed by Demina and Kudryashov to calculate the exact meromorphic solution. An important aspect of this analysis is to find Laurent series. This series will be useful for constructing the right meromorphic solution, and simultaneously proving the existence or nonexistence of the meromorphic solution \cite{Eremenko_Mero_KS_1}. At the end of this paper, we show the solitary wave solution as a special case of the modified quintic complex Ginzburg-Landau equation by selecting the condition of the physical system parameters. We show this solution to verify the solution shown in \cite{Kengne2007}. The structure of this paper is as follows. In Section 2, we describe the process of transforming the equation (\ref{Pers_GL}) to a nonlinear differential equation. The main part is Section 3, where we analyze the solution of the meromorphic equations using the \textit{Demina-Kudryashov} algorithm.

\section{Transformation Equation (\ref{Pers_GL}) to Nonlinear Ordinary Differential Equation}
In this section, we will show briefly the transformation of the equation (\ref{Pers_GL}) to nonlinear ordinary differential equation \cite{Kengne2006}. Firstly, we take the form
\begin{equation}
\label{psi_sd_a_expiphi} \psi(x,t)=a(x,t)\exp (i \varphi(x,t)),
\end{equation}
where $a(x,t)$ and $\varphi(x,t)$ are real. Inserting equation (\ref{psi_sd_a_expiphi}) into equation (\ref{Pers_GL}) and then separating its real  and imaginary parts to obtain
\begin{eqnarray}
\label{Pers_GL_riil} a \frac{\partial \varphi}{\partial t}-Pa\left(\frac{\partial \varphi}{\partial x}\right)^2+P\frac{\partial^2a}{\partial x^2}+\gamma a+Q_1a^5\nonumber \\
-Q_2a^3\frac{\partial \varphi}{\partial x}+Q_3a^3\frac{\partial \varphi}{\partial x}= 0,\\
\label{Pers_GL_imajiner} -\frac{\partial a}{\partial t}+2P\frac{\partial \varphi}{\partial x}\frac{\partial a}{\partial x}+Pa\frac{\partial^2\varphi}{\partial x^2}+Q_2a^2\frac{\partial a}{\partial x}=0.
\end{eqnarray}
Then, by defining (based on traveling wave model)
\begin{eqnarray}
\label{Model_TW_a} a(x,t)&=& a_0+\alpha(x-vt)=a_0+\alpha(z),\\
\label{Model_TW_phi} \varphi(x,t)&=&\phi(z) -(q_0-l_0v)t,
\end{eqnarray}
where $a_0, l_0, q_0$ and $v$ are real constants; $x-vt=z$. Then, substituting equation (\ref{Model_TW_a}) and (\ref{Model_TW_phi}) into equation (\ref{Pers_GL_riil}) and (\ref{Pers_GL_imajiner}) yields
\begin{eqnarray}
\label{Pers_GL_Trans_1} a(\gamma-q_0+l_0v)+P\left(\frac{d^2\alpha}{dz^2}-a\left(\frac{d\phi}{dz}\right)^2\right)\nonumber\\
+Q_1a^5+\left((Q_3-Q_2)a^2-v\right)a\frac{d \phi}{dz}=0
\end{eqnarray}
and
\begin{equation}
\label{Pers_GL_Trans_2} P\left(2\frac{d\phi}{dz}\frac{d\alpha}{dz}+a\frac{d^2\phi}{dz^2}\right)+\left((Q_2+Q_3)a^2+v\right)\frac{d\alpha}{dz}=0.
\end{equation}
Multiplying equation (\ref{Pers_GL_Trans_2}) with $a$ and then integrating it yields
\begin{equation}
\label{Pers_GL_Trans_3} \frac{d\phi}{dz}=\frac{K_1}{Pa^2}- \frac{Q_2+Q_3}{4P}a^2-\frac{v}{2P}
\end{equation}
where $K_1$ is the integration constant. Inserting equation (\ref{Pers_GL_Trans_3}) into equation (\ref{Pers_GL_Trans_1})  to obtain
\begin{eqnarray}
\frac{d^2a}{dz^2}&=& \frac{1}{4P^2}\left(\begin{array}{c}
-v^2-4P(\gamma-q_0+l_0v)\\-2K_1(Q_2+Q_3)
\end{array}\right)a \nonumber\\
\label{Pers_GL_Trans_4}&+& 4K_1(Q_2-Q_3)+\frac{v(Q_3-Q_2)}{2P^2}a^3\nonumber\\
&+&\frac{(5Q_3-3Q_2)(Q_2+Q_3)-16PQ_1}{16P^2}a^5\nonumber\\&+&\frac{K_1^2}{P^2a^3}.
\end{eqnarray}
Multiplying equation (\ref{Pers_GL_Trans_4}) by  $\frac{da}{dz}$ and integrating to obtain
\begin{eqnarray}
\left(\frac{da}{dz}\right)^2&=&\frac{1}{4P^2}\left(\begin{array}{c}
-v^2-4P(\gamma-q_0+l_0v)-\\ 2K_1(Q_2+Q_3)+4K_1(Q_2-Q_3)
\end{array}\right)a^2 \nonumber\\
&+& \frac{(5Q_3-3Q_2)(Q_2+Q_3)-16PQ_1}{48P^2}a^6 \nonumber \\
\label{Pers_GL_Trans_5}&+&\frac{v(Q_3-Q_2)}{4P^2}a^4- \frac{K_1^2}{P^2a^2}+\frac{K_2}{4}
\end{eqnarray}
where $K_2$ is integration constant. Then, using $a^2=w$, we obtain the following form
\begin{equation}
\label{Pers_GL_PDB_eliptik} \left(\frac{dw}{dz}\right)^2=-\frac{4K_1^2}{P^2}+K_2w+Cw^2+Dw^3+Ew^4
\end{equation}
where
\begin{eqnarray*}
	C&=&\frac{-v^2-4P(\gamma-q_0+l_0v)+2K_1(Q_2-3Q_3)}{P^2},\\
	D&=&\frac{v(Q_3-Q_2)}{P^2,}\\
	E&=&\frac{(5Q_3-3Q_2)(Q_2+Q_3)-16PQ_1}{12P^2}.
\end{eqnarray*}
Equation (\ref{Pers_GL_PDB_eliptik}) is the main equation solved in this paper. In the following sections, we will construct the meromorphic solutions of this equation using the \textit{Demina-Kudryashov} algorithm.

\section{Meromorphic Solutions of Nonlinear Ordinary Differential Equation }
In this section we will find the Laurent series and the meromorphic solutions of equation (\ref{Pers_GL_PDB_eliptik}). Firstly, by inserting  \cite{Demina_Kud_Explicit_Expression,Demina_Kud_From_Lau_Kawa,Kud_Mer_Sol_N_ODE,Eremenko_Mero_KS_1,Eremenko_Mero_KS_2}
\begin{equation}
\label{Pers_Utama_Uraian_Laurent_Paper} w(z)=\sum_{k=m}^{\infty}c_k(z-z_0)^k, m<0, c_m\neq 0
\end{equation}
into equation (\ref{Pers_GL_PDB_eliptik}), without loosing the generality, and not taking  $z_0$ into account, we find two kinds of Laurent series as follows
\begin{eqnarray}
\label{Sol_Laurent_GL_jenis1} w^{(1)}(z)=\frac{1}{\sqrt{E}z}-\frac{D}{4E}+\frac{1}{\sqrt{E}}\left(\frac{D^2}{16E}-\frac{C}{6}\right)z\nonumber \\
+\left(\frac{DC}{16E}-\frac{K_2}{8}-\frac{D^3}{64E^2}\right)z^2+....
\end{eqnarray}
and
\begin{eqnarray}
\label{Sol_Laurent_GL_jenis2} w^{(2)}(z)=-\frac{1}{\sqrt{E}z}-\frac{D}{4E}-\frac{1}{\sqrt{E}}\left(\frac{D^2}{16E}-\frac{C}{6}\right)z\nonumber \\
+\left(\frac{DC}{16E}-\frac{K_2}{8}-\frac{D^3}{64E^2}\right)z^2+....
\end{eqnarray}

As prevailed on power equations, we can see that the equation (\ref{Sol_Laurent_GL_jenis1}) and (\ref{Sol_Laurent_GL_jenis2})  are satisfied if $E \neq 0$. This is one of many necessary conditions of equation solutions that we need.

At a glance we see that the $K_1$ integration constant does not appear in the equation (\ref{Sol_Laurent_GL_jenis1}) and (\ref{Sol_Laurent_GL_jenis2}). But, actually $K_1$ is contained in $C$ constant and coefficient $c_3$, and other high order coefficients.

The Laurent series solution (\ref{Sol_Laurent_GL_jenis1}) and (\ref{Sol_Laurent_GL_jenis2}) have a simple pole. We can see from the Laurent series that the total residue is zero. This is the necessary existence condition to construct an elliptic equation solution (\ref{Pers_GL_PDB_eliptik}).


\subsection{Simply periodic solution to equation (\ref{Pers_GL_PDB_eliptik})}
Under the necessary existence condition of the solution, the Demina-Kudryashov algorithm can not be used to construct elliptic solutions for the first type. Therefore, we need to construct a simply periodic solution as follows. First, based on
\cite{Demina_Kud_Explicit_Expression,Demina_Kud_From_Lau_Kawa}, we have
\begin{equation}
\label{Sol_SP_GL_T1_awal} w(z)=\sqrt{L}c_{-1}^{(1)}\cot\left(\sqrt{L}z\right)+h_0.
\end{equation}
Expanding equation (\ref{Sol_SP_GL_T1_awal}) around $z=0$ yields
\begin{equation}
\label{Sol_SP_GL_T1_eksp_z0} w(z)=\frac{c_{-1}^{(1)}}{z}+h_0-\frac{c_{-1}^{(1)}L}{3}z-\frac{c_{-1}^{(1)}L^2}{45}z^3-...\qquad .
\end{equation}
Comparing (\ref{Sol_SP_GL_T1_eksp_z0}) and (\ref{Sol_Laurent_GL_jenis1}), we found
\begin{eqnarray*}
	c_{-1}^{(1)}=\frac{1}{\sqrt{E}}; \: h_0&=&-\frac{D}{4E}; \: L=\frac{C}{2}-\frac{3D^2}{16E};\\
	K_2&=&\frac{4CDE-D^3}{8E^2};
\end{eqnarray*}
\begin{eqnarray}
\label{Konstanta_SP_GL} K_1=\pm\frac{1}{16} \sqrt{\frac{8CD^2P^2}{E^2} - \frac{D^4P^2}{E^3} - \frac{16 C^2 P^2}{E}}.
\end{eqnarray}
So, simply periodic solution for equation (\ref{Pers_GL_PDB_eliptik}) is
\begin{equation}
\label{Sol_SP_GL_T1_akhir}  w(z)=\sqrt{\frac{C}{2E}-\frac{3D^2}{16E^2}}\cot\left(\sqrt{\frac{C}{2}-\frac{3D^2}{16E}}z\right)-\frac{D}{4E},
\end{equation}
with the parameter relation $C, D$ and  $E$ being 
\begin{equation}
\label{Relasi_E_D_C} 
E=\frac{9 D^2}{16 C}.
\end{equation}

The equation (\ref{Sol_SP_GL_T1_akhir}) is a simply periodic solution of the equation (\ref{Pers_GL_PDB_eliptik}), from which other solutions can be obtained using trigonometric and hyperbolic identities, such as soliton solutions or other solutions that differ only periodically according to their trigonometric identities. We can verify it by inserting the equation (\ref{Sol_SP_GL_T1_akhir}) into the equation (\ref{Pers_GL_PDB_eliptik}), then using relation parameters (\ref{Relasi_E_D_C}) and constants (\ref{Konstanta_SP_GL}), then the equation (\ref{Sol_SP_GL_T1_akhir}) satisfies the equation (\ref{Pers_GL_PDB_eliptik}). This clearly proves that the solutions we produce are true and verifiable.

\subsection{Rational solution}
Rational solution takes the form
\begin{equation}
\label{Sol_rasional_GL} w(z)=\pm\frac{1}{\sqrt{E}z}-\frac{D}{4E},
\end{equation}
with constants
\begin{eqnarray}
\label{konstanta_rasional}
K_2&=&\frac{4CDE-D^3}{8E^2},\nonumber\\
K_1&=& \pm\frac{i\sqrt{D}\sqrt{3D^3-16CDE+64E^2K_2}P}{32E^{3/2}},
\end{eqnarray}
and the parameter relation $C, D,$ and $E$
\begin{equation}
\label{parameter_rasional}
E=\frac{3D^2}{8C}.
\end{equation}

The two rational solutions (\ref{Sol_rasional_GL}) that we obtain (using these constants (\ref{konstanta_rasional}) and the parameter relation (\ref{parameter_rasional}) has also been inserted into equation (\ref{Pers_GL_PDB_eliptik}) and the results have been found to satisfy the equation. Again, this proves the truth of our solutions.

\subsection{Doubly periodic solution}
Now, we construct elliptic solution (doubly periodic) for second type. Based on \cite{Demina_Kud_Explicit_Expression,Demina_Kud_From_Lau_Kawa}, we can write
\begin{equation}
\label{Sol_eliptik_GL_T2_awal}  w(z)=c_{-1}^{(1)}\zeta(z;g_2,g_3)+c_{-1}^{(2)}\zeta(z-a;g_2,g_3)+\tilde{h}_0,
\end{equation}
where $c_{-1}^{(1)}=-c_{-1}^{(2)}= 1/\sqrt{E}$. Expanding equation (\ref{Sol_eliptik_GL_T2_awal}) around $z=0$, yields
\begin{eqnarray}
\label{Sol_eliptik_GL_T2_eksp_z0} w(z)=\frac{c_{-1}^{(1)}}{z}+h_0-c_{-1}^{(2)}Az+\frac{c_{-1}^{(2)}}{2}Bz^2\nonumber\\
+c_{-1}^{(2)}\left(\frac{g_2-10A^2}{10}\right)z^3+...
\end{eqnarray}
Expanding equation (\ref{Sol_eliptik_GL_T2_awal}) around $z=a$, we found
\begin{eqnarray}
\label{Sol_eliptik_GL_T2_eksp_za} w(z)=\frac{c_{-1}^{(2)}}{z-a}&+&h_0-c_{-1}^{(1)}A(z-a)+\frac{c_{-1}^{(1)}}{2}B(z-a)^2\nonumber\\
&+& c_{-1}^{(1)}\left(\frac{g_2-10A^2}{10}\right)(z-a)^3+...
\end{eqnarray}
where $A\stackrel{def}{=}\wp(a)$, $B\stackrel{def}{=}\wp_z(a)$, $\wp_z=d \wp/{dz}$; $h_0\stackrel{def}{=} \tilde{h_0}-c_{-1}^{(2)}\zeta(a)$; $\wp\equiv \wp(z,g_2,g_3)$ is  Weierstrass-$\wp$ elliptic function and $\zeta$ is the Weierstrass-$\zeta$ elliptic function; and $z=a$ is a pole of the second order type. Invariants $g_2$ and $g_3$ are determined from elliptic function $\wp$, and take the form
\begin{equation}
\label{g2g3} g_2=\sum_{\omega\neq 0}^{}\frac{60}{\omega^4}; g_3=\sum_{\omega\neq 0}^{}\frac{140}{\omega^6}
\end{equation}
where $\wp$, $g_2$, and $g_3$ satisfy the equation
\begin{equation}
\label{Pers_invarian_g2g3} \left(\wp_z\right)^2=4\wp^3-g_2\wp-g_3.
\end{equation}
We can rewrite equation (\ref{Sol_eliptik_GL_T2_awal}) as
\begin{equation}
\label{Pers_Sol_eliptik_awal_rewritten} w(z)=\frac{c_{-1}^{(2)}(\wp_z+B)}{2(\wp-A)}+h_0.
\end{equation}
Comparing equation (\ref{Sol_eliptik_GL_T2_eksp_z0}) and (\ref{Sol_Laurent_GL_jenis1}) and then equation (\ref{Sol_eliptik_GL_T2_eksp_za}) with (\ref{Sol_Laurent_GL_jenis2}), yield
\begin{eqnarray*}
	c_{-1}^{(1)}&=&-c_{-1}^{(2)}=\frac{1}{\sqrt{E}}; \: h_0=-\frac{D}{4E}; \: A=\frac{D^2}{16E}-\frac{C}{6},\\
	B&=&\sqrt{E}\left(\frac{K_2}{4}+\frac{D^3}{32E^2}-\frac{CD}{8E}\right),\\
	g_2&=&\frac{17 C^2}{36} + \frac{5D^4}{64E^2} - \frac{5CD^2}{12E}+\frac{DK_2}{4}+\frac{4EK_1^2}{P^2},
\end{eqnarray*}
\begin{eqnarray}
\label{Konstanta_solusi_eliptik}
g_3=\frac{11 C^3}{108} &+& \frac{7 C^2 D^2}{192 E}+ C\left(\frac{7 g_2}{6}+\frac{5 D K_2}{6}+\frac{4 E K_1^2}{P^2}\right)\nonumber\\
&+&\frac{1}{64} \left(\begin{array}{c}
-\frac{7 D^3 K_2}{E}- 4 E K_2^2\\ + D^2\left(-\frac{28 g_2}{E}-\frac{96 K_1^2}{P^2}\right)
\end{array}\right).
\end{eqnarray}
So, the elliptic solution for second type is
\begin{equation}
\label{Sol_eliptik_GL_T2_akhir} w(z)=\frac{6 (4 \sqrt{E} \frac{d\wp}{dz}+ 2 D \wp+E K_2)-C D }{3 D^2 - 8 E (C + 6 \wp)}.
\end{equation}

The equation (\ref{Sol_eliptik_GL_T2_akhir}) is a doubly periodic explicit solution of the equation (\ref{Pers_GL}). We can clearly see that this solution is more explicit than that shown in \cite{Kengne2007}, since based on this algorithm we do not need to define the "new $R (w)$" function on the right side of the nonlinear differential equation (\ref{Pers_GL}). An important aspect is as we did before in simply periodic solutions and rational solutions; the obtained solution is then inserted into equation (\ref{Pers_GL_PDB_eliptik}), and the result is found to satisfy the equation (of course by using constants and parameter relations (\ref{Konstanta_solusi_eliptik})). Thus, the more reinforcing that our solutions are true. We will show that by choosing a special case, we can find a solitary wave kink solution as shown in \cite{Kengne2007}.

Then, if we try to find a second simply periodic solution of the Ginzburg-Landau equation, we can find constants like $L_j$ and $A_j$ with $ j = 1,2,3 $ and integration constants $K_1$ and $K_2$, and of course we can find a simply periodic solution form. But the problem is how to find the relationship parameters of other constants like $C, D, E$ and also their relations to the integration constant. This is because, based on the two Laurent series above, even for this series, we always find the same value for both series. Therefore, we can not find the relationship of these constants and hence we can not verify the solution for the equation (\ref{Pers_GL_PDB_eliptik}). This means that the solution (\ref{Sol_SP_GL_T1_akhir}) is enough for us to construct other types of solutions, especially to describe the traveling wave or solitary wave in the physical system.

\begin{figure}[!h]
	\centering
	\includegraphics[width=7cm]{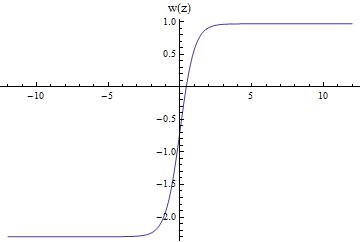}
	\caption{Kink solitary Solution for $C=3, D=1$ and $E=3/8$}
	\label{GambarKink}
\end{figure}

For example, we can construct a solitary wave solution. We have explained before that without losing the generality of the Laurent series form, we neglected the $z_0$ constant. However, in this part, we can rewrite the simply periodic solution (\ref{Sol_SP_GL_T1_awal}) containing $z_0$ as

\begin{equation}
\label{Sol_SP_GL_T1_awal_Kink}  w(z)=\sqrt{\frac{C}{2E}-\frac{3D^2}{16E^2}}\cot\left(\sqrt{\frac{C}{2}-\frac{3D^2}{16E}}(z-z_0)\right)-\frac{D}{4E}.
\end{equation}
We can choose $\frac{C}{2}-\frac{3D^2}{16E}<0$ to find cot hyperbolic function, then by using identity $\coth(z+i\pi/2)=\tanh z$, and $z_0=i\pi/2$, we find kink solitary wave solution as follows
\begin{equation}
\label{Sol_SP_GL_T1_akhir_Kink} w(z)=\sqrt{\frac{C}{2E}-\frac{3D^2}{16E^2}}\tanh\left(\sqrt{\frac{C}{2}-\frac{3D^2}{16E}}z\right)-\frac{D}{4E}.
\end{equation}
The solution (\ref{Sol_SP_GL_T1_akhir_Kink}) is a family kink solitary wave solution, shown in (\ref{GambarKink}). We can choose other values ​​from the above parameters to find other forms of solitary wave kink solutions.
Since we have chosen the condition $\frac {C} {2} - \frac {3D^2} {16 E} <0 $ to find the equation (\ref{Sol_SP_GL_T1_akhir_Kink}), this means that, when we want to choose a constant for visualize the solution (\ref{Sol_SP_GL_T1_akhir_Kink}), we must choose $ \frac {C} {2} - \frac {3D^2} {16 E}>0 $. Because, if we use the form above less than zero, it can return the hyperbolic form to the Tan form. This means that we can not find solitary wave solutions.

We visualize the solution as shown in Figure (\ref{GambarKink}) by choosing condition equal to 1, so we found the relation parameter $E = 3D ^ 2/8 (C-2)$; and for $C = 3, D = 1$ and $E = 3/8$, then generally for this case we can use $C\neq2$ and $C> 2$. In this case, we can not select a value for $C <2$, because it can generate $w (z)$ as imaginary. To find a solitary wave solution, we need $ w (z) $ real value. We can also choose the same condition with a higher value, and it can have an impact on the determination of the $ C $ constant. This solution is in accordance with the results shown by Nickel and Sch\"{u}rmann in \cite{Kengne2006} .They have found solitary kink wave solutions as the special case of this equation.

\section{Conclusion}
In this paper we solve the ordinary nonlinear differential equations using algorithms proposed by Demina and Kudryashov. This equation is transformed from the modified quintic complex Ginzburg-Landau equation using the traveling wave model. We found two kinds of Laurent series solutions with simple poles and then we constructed meromorphic solutions. We find meromorphic solutions that contains simply periodic solutions, doubly periodic (elliptic) solutions and a rational solution. These solutions satisfy the equation (\ref{Pers_GL_PDB_eliptik}) with the parameter relation shown above.

For doubly periodic solutions, we find solutions in explicit form without having to define other variables. The solutions were more common than what Kengne and Liu invented \cite{Kengne2006}. They find the solution only for some special cases. This meromorphic solution is a family of solutions to construct many other solutions we need to describe or solve the physical system.

We have shown a kink solitary wave  solution as an example or a special case of a meromorphic solution. This solution is in accordance with the results shown by Nickel and Sch\"{u}rmann in \cite{Kengne2007}. This particular solution is one of the other wave solutions that we can construct. We can construct many other solutions using trigonometric identities, hyperbolic functions or with other constants.

\section*{Acknowledgements}
\noindent 
HFL thanks Redi Pingak  for his constructive comments. The authors would like to thanks to ministry of research technology and higher education, Republic of Indonesia for supporting this work by providing grant Desentralisasi 2017, Research KK ITB 2017. WH would like to thank to ministry of research technology and higher education, Republic of Indonesia for PhD Scholarship program.
\bibliographystyle{unsrt}
\bibliography{refs}

\begin{thebibliography}{10}

\bibitem{sulaeman2015}
H.~Alatas L. T.~Handoko A.~Sulaiman, F. P.~Zen.
\newblock Davydov’s soliton in an inhomogeneous medium.
\newblock {\em AIP Conf. Proc.}, 1656:050012, 2015.

\bibitem{sulaeman2012}
H.~Alatas L. T.~Handoko A.~Sulaiman, F. P.~Zen.
\newblock Dynamics of dna breathing in the peyrard-bishop model with damping
  and external force.
\newblock {\em Nonlinear Phenomena}, 241(9):1640--1647, 2012.

\bibitem{sulaeman2012b}
H.~Alatas L. T.~Handoko A.~Sulaiman, F. P.~Zen.
\newblock The thermal denaturation of the peyrardbishop model with an external
  potential.
\newblock {\em Physica Scripta}, 86(1):015802, 2012.

\bibitem{Kud_Mer_Sol_N_ODE}
N.~A. Kudryashov.
\newblock Meromorphic solutions of nonlinear ordinary differential equations.
\newblock {\em Communications in Nonlinear Science and Numerical Simulation},
  15:2778--2790, 2010.

\bibitem{Kud_Be_Careful_Exp_M}
N.A. Kudryashov and N.B Loguinova.
\newblock Be careful with exp-function method.
\newblock {\em Communications in Nonlinear Science and Numerical Simulation},
  14:1881--1890, 2009.

\bibitem{Kud_A_Note_on_Kawa_Exp}
N.A. Kudryashov.
\newblock A note on new exact solutions for the kawahara equation using
  exp-function method.
\newblock {\em Journal of Computational and Applied Mathematics},
  234:3511--3512, 2010.

\bibitem{Kud_Comenton_FisherEq_2009}
N.A. Kudryashov.
\newblock Comment on: A novel approach for solving the fisher equation using
  exp-function method.
\newblock {\em Physics Letters A}, 373:1196--1197, 2009.

\bibitem{Kud_Logui_Be_Careful_2009}
N.B.~Loguinova N.A.~Kudryashov.
\newblock Be careful with exp-function method.
\newblock {\em Communications in Nonlinear Science and Numerical Simulation},
  14:1881--1890, 2009.

\bibitem{Kud_Seven_Common_2009}
N.A. Kudryashov.
\newblock Seven common errors in finding exact solutions of nonlinear
  differential equations.
\newblock {\em Communications in Nonlinear Science and Numerical Simulation},
  14:3507--3523, 2009.

\bibitem{Kud_Sinelshchikov_A_Note_JMeq_2010}
N.A. Kudryashov and D.~I. Sinelshchikov.
\newblock A note on abundant new exact solutions for the (3+1)-dimensional
  jimbo-miwa equation.
\newblock {\em Journal of Mathematical Analysis and Applications},
  371:393--–396, 2010.

\bibitem{Kud_n_Sinel_A_Note_FifthO_KdV_Exp}
N.A. Kudryashov and N.B Loguinova.
\newblock A note on ”exp-function method for the exact solutions of fifth
  order kdv equation and modified burgers equation”.
\newblock {\em Journal of Mathematical Analysis and Applications}, 2010.

\bibitem{Park_A_Note_on_Seventh_o_KdV}
E.J. Parkes.
\newblock A note on travelling - wave solutions to lax’s seventh - order kdv
  equation.
\newblock {\em Appl. Math. Comput}, 215:864--865, 2009.

\bibitem{Assas}
Assas L.M.B.
\newblock New exact solutions for the kawahara equation using exp-function
  method.
\newblock {\em Journal of Computational and Applied Mathematics}, 233:97--102,
  2009.

\bibitem{Zheng2009}
Shan W.~R Zheng~Z.
\newblock Application of exp-function method to the whitham - broer - kaup
  shallow water model using symbolic computation.
\newblock {\em Applied Mathematics and Computation}, 215:2390--2396, 2009.

\bibitem{Salas}
Hernandez~J.E.C. Salas~A.H.S., Gomez~C.A.S.
\newblock New abundant solutions for the burgers equation.
\newblock {\em Computers and Mathematics with Applications}, 58:514--520, 2009.

\bibitem{Demina_Kud_Explicit_Expression}
Maria~V. Demina and N.~A. Kudryashov.
\newblock Explicit expressions for meromorphic solutions of autonomous
  nonlinear ordinary differential equations.
\newblock {\em Commun Nonlinear Sci Numer Simulat}, 16:1127--–1134, 2011.

\bibitem{Demina_Kud_From_Lau_Kawa}
Maria~V. Demina and N.~A. Kudryashov.
\newblock From laurent series to exact meromorphic solutions: The kawahara
  equation.
\newblock {\em Physics Letters A}, 374:4023–--4029, 2010.

\bibitem{Kengne2006}
E.~Kengne and W.~M. Liu.
\newblock Exact solutions of the derivative nonlinear schrödinger equation for
  a nonlinear transmission line.
\newblock {\em Phys. Rev. E}, 73:026603--1--026603--8, 2006.

\bibitem{Kengne2007}
J.~Nickel and H.~W. Schurmann.
\newblock Comment on “exact solutions of the derivative nonlinear
  schrödinger equation for a nonlinear transmission line”.
\newblock {\em Phys. Rev. E}, 75, 2007.

\bibitem{Whittaker_Whatson_A_Course}
W.~T. Whittaker and G.~N. Watson.
\newblock {\em A Course of Modern Analysis}.
\newblock Cambridge Mathematical Library, 1927.

\bibitem{Eremenko_Mero_KS_1}
A.~Eremenko.
\newblock Meromorphic traveling wave solutions of the kuramoto-sivashinsky
  equation.
\newblock {\em arXiv:nlin.SI/0504053 v1 25 Apr 2005}, 2005.

\bibitem{Eremenko_Mero_KS_2}
A.~Eremenko.
\newblock Meromorphic traveling wave solutions of the kuramoto-sivashinsky
  equation.
\newblock {\em Journal of Mathematical Physics, Analysis, Geometry},
  2:278--286, 2006.

\end{thebibliography}

\end{document}